\documentstyle[11pt,paspconf]{article}

\def\apj#1{{ ApJ,} { #1}}
\def\mn#1{{ MNRAS,} { #1}}

\def\nat#1{{ Nature,} { #1}}
\def\apjs#1{{ ApJS,} { #1}}
\def\etal{{ et al.}\ }
\def\Mpc{$h^{-1}$~{\rm Mpc}}

\begin{document}

\title{Dark Matter in Groups and Clusters of Galaxies}

\author{Jaan Einasto and Maret Einasto}

\affil{Tartu Observatory, Estonia}

\begin{abstract}

We compare the characteristics of stellar populations with those of
dark halos.  Dark matter around galaxies, and in groups, clusters and
voids is discussed. Modern data suggest that the overall density of
matter in the Universe is $\Omega_M = 0.3 \pm 0.1$, about $80$~\% of
this matter is non-baryonic dark matter, and about $20$~\% is
baryonic, mostly in the form of hot intra-cluster and intragroup gas,
the rest in stellar populations of galaxies.  All bright galaxies are
surrounded by dark matter halos of external radii $200 - 300$~kpc;
halos consist mostly of non-baryonic matter with some mixture of hot
gas. The Universe is dominated by dark energy (cosmological constant)
term. Dark matter dominates in the dynamical evolution of galaxies in
groups and clusters.

\end{abstract}

\keywords{dark matter, galaxies, groups of galaxies, clusters of
galaxies, observations }

\section{History of the dark matter concept}

The story of dark matter is a classical example of a scientific
revolution (Kuhn 1970, Tremaine 1987).  It is impossible in this
review talk to discuss all aspects of dark matter.  We start with a
historical introduction, followed by a comparison of ordinary
stellar populations and the nature of dark matter.  Thereafter we
consider dark matter in galaxies, groups and clusters of galaxies, and
in voids; we also discuss the mean density of matter in the Universe.

First hints on the presence of a mass paradox in galaxies and clusters
of galaxies came over 60 years ago.  Oort (1932) noticed that there
may exist a discrepancy between the dynamical estimate of the local
density of matter in the Solar neighborhood in the Galaxy, and the
density of luminous matter.  Known stellar populations may be
insufficient to explain the vertical gravitational attraction in the
Galaxy which causes motions of stars perpendicular to the plane of the
Galaxy.  Zwicky (1933) measured radial velocities of galaxies in the
Coma cluster and found that the mass of the cluster exceeds the summed
mass of its galaxies more than tenfolds.  These studies raise two
problems, the one of the local dark matter in the disk of the Galaxy,
and the global dark matter penetrating clusters of galaxies.  In the
1930s astronomers were very busy to understand the evolution of stars,
and dark matter problems escaped the attention of the astronomical
community.

The next essential step in the dark matter story was made by Kahn \&
Woltjer (1959).  They noticed that the Andromeda galaxy and our Galaxy
approach each other, whereas almost all other galaxies recede from
us. The total mass of the Local Group, inferred from ascribing the
approach velocity to mutual attraction, exceeds the conventional mass
of M31 and Galaxy approximately tenfold.  This discovery again did not
attract much attention.  During the discussion of the stability of
clusters of galaxies in the 1960s, Ambartsumian suggested an opposite
view that clusters may be recently formed and expanding systems. This
suggestion contradicts, however, data on ages of cluster galaxies, see
van den Bergh (1999).

In the late 1960s and early 1970s it was realized that mass paradox
may be a global problem for all bright galaxies. Einasto (1969),
Sizikov (1969) and Freeman (1970) noticed that rotation velocities of
galaxies decrease more slowly in the outskirts of galaxies than
expected from the distribution of light.  Two possibilities were
discussed to explain this discrepancy -- systematic deviations from
circular motion or the presence of some massive but invisible
population in the outskirts of galaxies.  

One approach that has led to the conclusion of the presence of dark
matter around galaxies was the modeling of galaxies using a
combination of all available observational data on stellar populations
in galaxies of different morphological type. Such combined models were
reported during the First European Astronomy Meeting in Athens in
September 1972 (Einasto 1974). It was shown that ordinary stellar
populations cannot explain almost flat rotation curves of the outer
parts of spiral galaxies.  To explain flat rotation curves the
presence of a new invisible population, a ``dark corona'', was
suggested.  Independent evidence for the presence of dark matter
around galaxies was inferred by Ostriker and Peebles (1973) based on
disk stability arguments.  Available data were, however, not
sufficient to determine the total mass and dimension of the
hypothetical dark population.

To derive the mass distribution at larger distances from galactic
centers the teams at Tartu and Princeton investigated the dynamics of
companions of bright galaxies. They demonstrated that the internal
mass, inferred from the motion of companion galaxies, increases with
distance from centers of bright galaxies up to several hundred
kiloparsec, thus increasing the hitherto assumed dimensions and masses
of galaxies by an order of magnitude (Einasto, Kaasik \& Saar 1974,
Ostriker, Peebles \& Yahil 1974).  These studies suggest that the
presence of dark matter is a general property of galaxies and systems
of galaxies; this matter has a dominant contribution to the mass
budget in the Universe.  Difficulties connected with this
interpretation of rotation curves and dynamics of companion galaxies
were discussed by Burbidge (1975).  These three studies triggered the
boom of the dark matter studies.

Dark matter was discussed during the Third European Astronomical
Meeting in Tbilisi, in July 1975.  This Meeting was the highlight of
the dark matter discussion where supporters (Bertola \& Tullio 1976,
Einasto et al. 1976) and opponents (Karachentsev 1976, Oleak 1976,
Materne \& Tammann 1976, Fesenko 1976) of the concept of dark matter
had a hot debate.  The majority of speakers argued against the dark
matter concept; in the summary of the Meeting Kharadze noticed that
the dark matter concept did not find support.

The next public discussion of the dark matter problem was during the
IAU General Assembly in Grenoble in August 1976.  Here the focus was
the nature of the dark population.  Ostriker, Peebles \& Yahil (1974)
assumed that dark halos consist of faint stars; this concept was
discussed by Maarten Schmidt.  Population studies led the group at
Tartu to conclude that dark matter cannot be made of ordinary stars
but must have a different origin (Jaaniste \& Saar 1975). To make a
clear distinction between known halo population (which consists of old
stars) and the new population the term ``corona'' was suggested
(Einasto 1974).  The difference between ordinary galactic populations
and the dark matter population was summarized by Einasto, J\~oeveer \&
Kaasik (1976).  Ivan King from the audience noticed ``perhaps really
there are two halos in galaxies, stellar and dark''.  Initially hot
gas was considered as a possible candidate for the dark matter
(Einasto 1974). However, subsequent studies by Komberg \& Novikov
(1975), and Chernin et al. (1976) demonstrated that only a fraction of
the corona may be gaseous.  X-ray observations have confirmed that the
mass of hot gas in coronae is comparable with the mass of stellar
populations, however, hot gas is not sufficient to explain the total
``missing'' mass.  Thus the origin of coronae remained unclear.

The final acceptance of the presence of dark matter around galaxies
came after Morton Roberts, Vera Rubin and their collaborators had
shown that the outer parts of practically all spiral galaxies have
flat rotational curves (Roberts \& Whitehurst 1975, Rubin, Ford \&
Thonnard 1978, 1980, Rubin 1987).  However, theorists accepted the
presence of dark matter only after its role in the evolution of the
structure of the Universe was realized. This illustrates the
Eddington's test: ``No experimental result should be believed until
confirmed by theory'' (Turner 1999b). It was clear that, if nature
created so much dark matter, it must have some purpose.  Rees (1977)
noticed that neutrinos can be considered as dark matter particles; and
Chernin (1981) showed that, if dark matter is non-baryonic, then this
helps to explain the paradox of small temperature fluctuations of
background microwave radiation.  Density perturbations of non-baryonic
dark matter start growing already during the radiation-dominated era
whereas the growth of baryonic matter is damped by radiation.  If
non-baryonic dark matter is dynamically dominating, the total density
perturbation can have an amplitude of the order $10^{-3}$ at the
recombination epoch, which is needed for the formation of the observed
structure of the Universe.  Baryonic matter flows after recombination
to gravitational wells formed by non-baryonic matter.  Chernin
considered neutrinos with non-zero rest mass as a possible candidate,
but other non-baryonic particles do the job as well. This result was
discussed in a conference in Tallinn in spring 1981.  In the summary
speech of this conference Zeldovich concluded: ``Observers work hard
to collect data, theorist interpret observations; are often in error,
correct their errors and try again; and there are only very rare
moments of clarification.  Today it is one of such rare moments when
we have holy feeling of understanding Nature. Non-baryonic dark matter
is needed to start structure formation early enough''.

Soon it was realized that neutrino-dominated or hot dark matter
generates almost no fine structure of the Universe -- galaxy filaments
in superclusters (Zeldovich, Einasto \& Shandarin 1982), and that the
structure forms too late (White, Davis \& Frenk 1984).  A much better
candidate for dark matter is some sort of cold particles as axions
(Blumenthal et al. 1984).  The dark matter concept as a solid basis of
the contemporary cosmology was incorporated in full details in a
series of lectures by Primack (1984), and was discussed in the IAU
Symposium on Dark Matter (Kormendy \& Knapp 1987).  Thus in the end it
took over fifty years from the first discoveries by Oort and Zwicky
until the new paradigm was generally accepted.  However, the story is
not over.  The nature of the dark matter is still unclear -- we do not
know exactly what the cold dark matter is, and whether it is mixed
with hot dark matter (neutrinos).

\section{Galactic populations}

Dark matter is invisible, and the only possibility to determine its
mass, radius and shape in galaxies and clusters of galaxies is
modeling of populations present in these systems, using all available
observational data on the distribution of populations and on dynamics
of the system.  Thus our first task is to find the main parameters of
stellar populations in galaxies.

Models of stellar populations in galaxies were constructed by Einasto
(1974), and more recently by Einasto \& Haud (1989), Bertola et
al. (1993) and Persic, Salucci \& Stel (1996), among others. Models
use luminosity profiles of galaxies, rotation curves, velocity
dispersions of central stellar clusters, and other relevant data.
Parameters can be determined for the stellar halo, the bulge and the
disk, and for the dark population.  To determine the amount of dark
matter in and around galaxies, the mass-to-luminosity ratio of the
stellar population, $M/L_B$, is of prime importance. The available
data show that the mean $M/L_B$ is surprisingly constant; for the
stellar halo it is of the order of unity, for the bulge approximately
3, and only for the metal-rich cores of massive galaxies the value
approaches 10.  The mean mass-to-luminosity ratio for all visible
matter, weighed with the luminosities of galaxies, is $M/L_B= 4.1 \pm
1.4$.  We can summarize properties of ordinary and dark populations as
follows:

1) Stellar populations have $1 \leq M/L_B \leq 10$, while dark
population has $M/L_B \gg 1000$.

2) There is a continuous transition of stellar populations from
stellar halo to bulge, from bulge to old disk, from old to young disk;
and intermediate populations are clearly seen in our Galaxy. All
stellar populations contain a continuous sequence of stars of
different mass, some of these stars have ages and masses which
correspond to ages and masses of luminous red giants, thus all stellar
populations are visible, in contrast to the dark population.

3) The density of stellar populations rapidly increases toward the
plane or the center of the galaxy, while the dark matter population
shows a much lower concentration of mass to the galactic plane and
center.

These arguments show quantitatively that dark matter must have an
origin different from stellar populations.  Since old and young
stellar populations form a continuous sequence, the dark population
must have originated much earlier.  There must be a large gap between
the formation time of dark halo and oldest visible stellar
populations, since there appears to be no intermediate populations
between the dark and stellar populations (Einasto, J\~oeveer \& Kaasik
1976).

\section{Dark matter in galaxies}

The possible existence of dark matter near the plane of the Galaxy was
advocated by Oort (1932, 1960), who determined the density of matter
in the Solar vicinity and found that there may be a discrepancy
between the dynamical density and the density calculated from the sum
of densities of known stellar populations.  This discrepancy was
studied by Kuzmin (1952), Eelsalu (1959) and J\~oeveer (1972, 1974);
all three independent analyses demonstrated that there is practically
no local mass discrepancy in the Galaxy.  A much higher value of the
local dynamical density was found by Bahcall (1984a, 1984b, 1987).  It
is clear that non-baryonic dark matter cannot contract to a flat
population needed to explain the presence of the local mass
discrepancy. For this reason it is natural to expect that the local
dark matter, if present, must be of stellar origin.  The local dark
matter problem has been analyzed by Gilmore (1990 and references
therein).  Most recent data suggest that dynamically determined local
density of mass is approximately $0.1~M_{\odot}~pc^{-3}$, in good
agreement with direct estimates of the density.  Thus there is no firm
evidence for the presence of local dark matter in our Galaxy.

Flat rotation curves of galaxies suggest that there must be another
dark population in galaxies. This global dark matter must have a
more-or-less spherical distribution to stabilize the flat population
(Ostriker \& Peebles 1973).  As discussed above the dark population
has properties completely different from properties of known stellar
populations.  It is generally believed that this population is
non-baryonic.

The mass and volume occupied by dark matter halos around galaxies can
be determined only on the basis of relative motions of visible objects
moving within these dark halos.  Almost all bright galaxies are
surrounded by dwarf companion galaxies, and in this respect they can
be considered as poor groups of galaxies (Einasto et al. 1974).  Such
clouds of satellites have radii 0.1 to 1~\Mpc.  The relative motions
of companion galaxies indicate that the total mass within the radius
of orbits grows approximately linearly with distance. This suggests
that dark halos of main galaxies have approximately isothermal density
profiles.  The outer radius of isothermal halos of giant galaxies is,
however, not well determined, since there are no objects which can
test the relative velocity at large distance from the main galaxy.
The Local Group of galaxies yields an unique possibility to measure
the relative radial velocity of two subgroups, located around our
Galaxy and the Andromeda galaxy.  These measurements show that the
total mass of the Local Group is $\approx 5 \times 10^{12}~M_{\odot}$
(Kahn \& Woltjer 1959, Einasto \& Lynden-Bell 1982).  Masses
determined from velocities of companions within both subgroups are $2
\times 10^{12}~M_{\odot}$ and $3 \times 10^{12}~M_{\odot}$, for our
Galaxy and M31, respectively.  In these determinations it is assumed
that dark halos of M31 and Galaxy have external radii about 200 --
300~kpc (Haud \& Einasto 1989, Tenjes, Haud \& Einasto 1994).  We see
that individual masses are in good agreement with to total mass
derived from the approach velocity; in other words this agreement
confirms that estimated external radii and masses of dark halos are
correct.

\section{Dark matter in clusters}

The distribution of mass in clusters of galaxies can be determined by
three independent methods: from the distribution of relative
velocities of galaxies, from the distribution and temperature of hot
X-ray emitting gas, and from the gravitational lensing effect.  All
three methods can be applied in the case of clusters of galaxies and
rich groups of galaxies, so we start our discussion from these
systems. 

\subsection{Rich clusters of galaxies}

The classical method to determine the mass distribution in clusters is
based on the measurements of the velocity dispersion of galaxies in
clusters.  The method may be biased since the number of clusters with
measured redshifts is usually small and it is difficult to exclude
foreground and background clusters, especially in regions of high
density of galaxies (superclusters).  During the last decade X-ray
measurements have supplied a more accurate method to determine masses
and mass profiles of clusters of galaxies.  The method is based on the
observation that hot gas and galaxies are in hydrostatic equilibrium
within a common cluster potential, i.e. both move under gravity in the
potential well of the cluster.  The mass distribution of the cluster
can be derived from the mean temperature of the gas and radial
gradients of the temperature and density (Watt et al. 1992, Mohr et
al. 1999).  The intensity of the X-ray emission gives information on
the mass distribution of hot gas, thus X-ray observations yield
simultaneously the distribution of the total mass and gas mass in the
cluster.  Galaxies give additional information on the distribution of
mass in galaxies, thus altogether three distributions can be found.
ROSAT X-ray satellite data are presently available for many clusters
and rich groups of galaxies.  As an example of the integrated mass
distribution in the Perseus cluster of galaxies we refer to
B\"ohringer (1995).  In other clusters studied so far the
distributions are rather similar.  The main conclusions from these
studies are the following:

1) the radial distributions of the total mass, gas mass, and galaxy
mass are similar;

2) intra-cluster hot gas constitutes $14 \pm 2$~\% of the total mass of
   clusters (for Hubble constant $h=0.65$);

3) the mass in visible populations of galaxies is $\approx 3$~\% of
   the total mass of the cluster.

X-ray data also yield the mass-to-luminosity ratio of the cluster:
$M/L_V = 150~h~M_{\odot}/V_{\odot}$ (David, Jones \& Forman 1996).
This mean value is valid for the whole range of temperatures and
masses of clusters and groups. Modern data based on velocity
dispersions of galaxies in clusters yield $M/L_V = 213 \pm
60~h~M_{\odot}/V_{\odot}$ (Carlberg et al. 1997).

Gravitational lensing yields another independent method to derive the
mass of clusters of galaxies.  This method has been applied for
several clusters, and the results are in agreement with masses
determined from X-ray data (Mellier, Fort \& Kneib 1993, Schindler et
al. 1995).

ROSAT data have been used to investigate the mass distribution in a
clusters filament in the core of the Shapley supercluster (Kull \&
B\"ohringer 1999).  Data show that there exist a continuous X-ray
emission along the filament joining three rich clusters of galaxies.
This emission indicates the presence of a potential well along the
filament filled with dark matter and hot gas.  A similar distribution
of galaxies along the main chain of the Perseus supercluster is known
long ago (J\~oeveer, Einasto \& Tago 1978).  These data indicate that
galaxies, hot gas and dark matter form similar condensations along
filaments joining clusters and groups of galaxies.

\subsection{Poor groups of galaxies}

Most galaxies in the Universe belong to poor groups with one or few
bright galaxies and a number of faint dwarf companions.  The Local
group is an example of poor groups with two major concentration centers.
The basic difficulty in the study of the mass distribution in poor
groups lies in the weakness of the X-ray emission and absence in most
groups bright companions to measure the relative velocity on large
distance from the groups center.

Available X-ray data suggest that poor groups have a lower fraction of
hot gas than do rich clusters of galaxies; the mass of hot gas is
approximately ten times smaller than the stellar mass (Ponman \&
Bertram 1993, Pildis, Bregman, \& Evrard 1995).  According to X-ray
data dark matter extends significantly beyond the apparent
configuration of bright galaxies in good agreement with optical data
on the distribution of faint companion galaxies; the total
mass-to-luminosity ratio is in agreement with optical data, $M/L_B
\approx 120~h~M_{\odot}/L_{\odot}$ (Ponman \& Bertram 1993).  Galaxies
in compact groups show signs of distortions which indicate that these
groups are formed as a result of orbital decay; galaxies merge within
a few billion years to form a giant elliptical galaxy in the center of
the group; this process is rather rapid.  The presence of such groups
indicates that there should exist fossil groups, consisting only of
the central giant elliptical galaxy surrounded by the dark matter of
the previous group.  Such fossil groups are actually observed,
examples are NGC~315 in the Perseus supercluster chain -- a massive
radio galaxy with very large radio lobes (J\~oeveer, Einasto \& Tago
1978), and NGC~1132 (Mulchaey \& Zabludoff 1999).

\subsection{Dynamics of main galaxies in groups and clusters}

If the dominant galaxies in groups were formed by merging of its
former companions one would expect the internal velocity dispersion of
these galaxies to be comparable with the velocity dispersion of
galaxies in the group before the merger event.  In Fig.~1 we plot the
velocity dispersion of central galaxies, $\sigma_{gal}$, in groups and
clusters as a function of the velocity dispersion of galaxies in
respective systems, $\sigma_{clust}$.  We see that the internal
velocity dispersion of dominant galaxies in rich clusters is much
lower than the velocity dispersion in clusters, but comparable to the
velocity dispersion of galaxies in subgroups often found in clusters.
This observation suggests that central galaxies formed already in the
early stages of cluster evolution, before subgroups merged with the
presently observed cluster.  A similar conclusion has been reached by
Dubinski (1998) using N-body simulations of cluster evolution.

\begin{figure*}[ht]
\vspace*{7cm} 
\caption{The relation between the velocity dispersion of the
dominant galaxy, $\sigma_{gal}$, and the velocity dispersion of the
host cluster, $\sigma_{clust}$. The relation is shown separately for
clusters with central cD galaxy and central E galaxy.  Straight line
marks the equality of both dispersions, $\sigma_{gal} = \sigma_{clust}$.
} 
\includegraphics{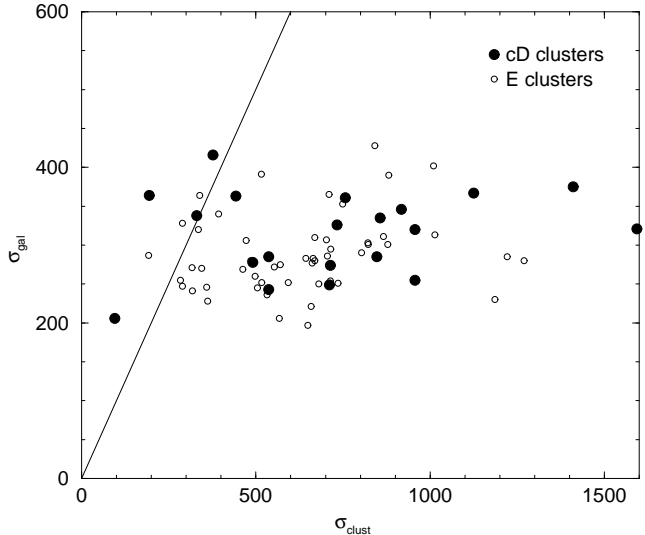}
\label{figure1}
\end{figure*}

\section{Dark matter in voids}

In the mid-1970s it was discovered that field galaxies are not
randomly distributed in space but form long filaments and chains
between clusters and groups; clusters and groups themselves are
concentrated to superclusters of galaxies (J\~oeveer \& Einasto 1978,
J\~oeveer, Einasto \& Tago 1978).  Between galaxy filaments there are
big volumes devoid of any visible form of matter.  One of the first
questions asked was: are these voids really empty or do they contain
some hidden matter?  This problem was investigated by Einasto,
J\~oeveer \& Saar (1980).  The study was based on the well-known
theory of the growth of density perturbations developed by Zeldovich
(1970) and Press \& Schechter (1974).  According to Zeldovich matter
flows away from under-dense regions towards high-density ones until
over-dense regions collapse and form galaxies.  The density of matter
in under-dense regions decreases approximately exponentially and never
reaches zero.  In order to form a galaxy or cluster the over-density
within a radius of $r$ must exceed a certain limit, about 1.68 in case
of spherical perturbations (Press-Schechter limit).  On the basis of
these considerations one can make two important conclusions: first,
there must exist some primordial matter in voids, and second, the
galaxy formation is a threshold phenomenon.  Recent hydrodynamical
simulations of the evolution of the density field and formation of
galaxies have confirmed these theoretical expectations (Cen \&
Ostriker 1992, 1999, Katz et al. 1992, 1996).

\begin{figure*}[ht]
\vspace*{7cm} 
\caption{The relation between the fraction of matter in galaxies,
$F_{gal}$, and $\sigma_8$. The thick bold solid line shows the definition
relation $(\sigma_8)_m = F_{gal} (\sigma_8)_{gal}$, while the bold
solid, the dashed and the dot-dashed lines give the relation obtained
from numerical simulations of how voids are emptied in different
cosmological models.  }  
\includegraphics{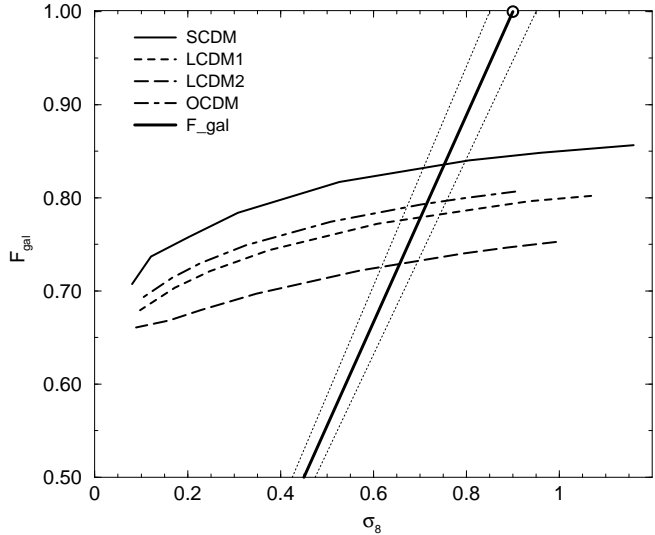}
\label{figure2}
\end{figure*}

Quantitative estimates of the total fraction of matter in voids have
been made by Einasto et al. (1994, 1999).  These estimates are based
on N-body calculations of the evolution of under- and over-density
regions for a variety of cosmological models.  The density field was
calculated using a small smoothing length, about 1~\Mpc, which
corresponds to the mean size of small groups of galaxies, dominant
structural elements of the Universe.  A problem in these calculations
is the identification of the present epoch of simulations.  This epoch
can be determined using $\sigma_8$ normalization of the density field.
The present mean value of density perturbations in a sphere of radius
8~\Mpc\ can be determined directly from observations.  Initially half
of all matter was located in regions below the mean density (this
follows from the simple fact that initial density perturbations are
very small).  During the evolution matter flows away from low-density
regions (see Fig.~2).  The present fraction of matter in voids is
somewhat model-dependent, $25 \pm 10$~\% of the total amount of matter
(Einasto et al. 1999).

\section{Mean density of matter in the Universe}

There are several independent methods to derive the mean density of
matter in the Universe.  The first method is based on primordial
nucleosynthesis data, which indicate that the baryon density is
$\Omega_{b} h^{2}= 0.019 \pm 0.002$ (Schramm \& Turner 1998, Turner
1999a). If we use a Hubble parameter of $h= 0.65 \pm 0.05$, and apply
the ratio of baryon to overall density as suggested by X-ray data, we
obtain for the mean density of matter $\Omega_M= 0.31(h/0.65)^{-1/2}
\pm 0.04$.  The second method uses mass-to-luminosity ratios of groups
and clusters, and the mean luminosity density.  This method gives the
density of the clustered matter. If we add the density of matter in
voids as suggested by void evacuation data, we get $\Omega_M= 0.25 \pm
0.05$ (Bahcall 1997, Einasto \etal 1999).  The distant supernova
project (Perlmutter et al. 1998, 1999, Riess et al. 1998) allows to
measure the curvature of the Universe and to distinguish between the
matter density, $\Omega_M$, and the cosmological constant parameter,
$\Omega_{\Lambda}$; this method suggests that the Universe is
dominated by the cosmological term, the density of matter is
$\Omega_M= 0.28 \pm 0.1$.  Similarly, the comoving maximum of the
galaxy power spectrum allows to measure the cosmological curvature
(Broadhurst \& Jaffe 1999), and prefers a Universe with $\Omega_M
= 0.4 \pm 0.1$.  As demonstrated by Bahcall \& Fan (1998) and
Eke \etal (1998), the rate of the evolution of cluster abundance
depends strongly on the mean density of the Universe.  The cluster
abundance method yields for the density a value $\Omega_M=0.3 \pm
0.1$.  Finally, the dynamics of the Local Group and its vicinity,
using the least action method, also yields a low density value (Shaya
et al. 1999).

The weighed mean of these independent methods is $\Omega_M= 0.30 \pm
0.05$; i.e. the overall density of matter in the Universe is
sub-critical by a wide margin. The quoted error is intrinsic, if we
add possible systematic  errors we get an error estimate $\pm 0.1$.
Supernova and CMB data exclude the possibility of an open Universe:
the dominating component in the Universe is the dark energy -- the
cosmological constant term or some other term with negative pressure
(Turner 1999b, Perlmutter et al. 1998, 1999).

\section{Summary}

The present knowledge of the dark matter in the Universe can be
summarized as follows.

1) The evidence for the presence of local dark matter in the disk of
   the Galaxy is not convincing; if present, it must be of baryonic
   origin as non-baryonic matter cannot form a flat disk.

2) The mean mass-to-luminosity ratio of stellar populations in
   galaxies is $M/L_B \approx 4~M_{\odot}/L_{\odot}$; the mean
   mass-to-luminosity ratio in groups and clusters of galaxies is $100
   - 200~h~M_{\odot}/L_{\odot}$.

3) The presence of dark matter halos of galaxies, and of dark common
   halos of groups and clusters is well established; the bulk of the
   dark population consists of some sort of cold dark matter. About $
   5$~\% of mass in poor groups, and $15$~\% in rich clusters is in
   the form of hot X-ray emitting gas.

4) There exists dark matter in voids; the fraction of matter in voids
   is $\approx 25$~\%, and in high-density regions $\approx 75$~\% of
   the total matter.

5) The total density of matter is, $\Omega_M =0.3 \pm 0.1$, and the
   density of dark energy (cosmological constant) is $\Omega_{\Lambda}
   = 0.7 \pm 0.1$.

6) On all scales larger than sizes of galaxies the dynamics
   is determined by  dark matter.

\acknowledgments The authors would like to thank H. Andernach for 
   suggestions on the presentation this work.

\end{document}